\documentstyle[amsfonts,aps,twocolumn]{revtex}
\begin{document}
\draft
\title{Duality in a fermion-like formulation for the electromagnetic field}
\author{Everton M. C. Abreu\thanks{%
Financially supported by Funda\c{c}\~ao de Amparo \`a Pesquisa do Estado de
S\~ao Paulo (FAPESP).} and Marcelo Hott}
\address{Departamento de F\'\i sica e Qu\'\i mica, Universidade Estadual Paulista,\\
Av. Ariberto Pereira da Cunha 333, Guaratinguet\'a, 12500-000, \\
S\~ao Paulo, SP, Brazil, \\
e-mail: everton@feg.unesp.br and hott@feg.unesp.br}
\date{\today}
\maketitle

\begin{abstract}
\noindent We employ the Dirac-like equation for the gauge field proposed by
Majorana to obtain an action that is symmetric under duality transformation.
We also use the equivalence between duality and chiral symmetry in this
fermion-like formulation to show how the Maxwell action can be seen as a
mass term.
\end{abstract}


%\begin{\large}

% o comando \draft colocas os pacs numbers

\section{Introduction}

Recently many papers have been published exploring the issue of duality.
This production has been motivated by the relation between duality symmetry
and theories with strong and weak coupling \cite{mo}. Another motivation has
been due to the presence of various duality symmetries in the string
theories \cite{fil}, specifically, the target space duality (T-duality),
which is the symmetry of the low energy effective field theory and the
S-duality (the generalization of the electric/magnetic duality) which is the
invariance under the $SL(2,R)$ duality transformations of the equations of
motion for the bosonic sector of the heterotic string. Besides these, a very
important interest resides in the electrically and magnetically charged
black holes in the semiclassical view \cite{DHT}.

Consequently, one of the first dualities observed, the electric-magnetic
duality in Maxwell's equations, has received great attention. However, the
problem that arises in preserving the duality symmetry and, at the same
time, the manifest Lorentz covariance when the duality is implemented is a
very difficult obstacle. The main objective in the literature is to
construct a duality symmetric action which is also manifestly Lorentz
invariant.

Schwarz and Sen (SS) \cite{SS} proposed a dual invariant action in which one
more potential has been introduced generalizing the T-duality symmetric
string action \cite{T} to the case of the heterotic string. Although this
formulation is not manifestly covariant it is classically and quantically
Poincar\'{e} invariant. One way to recover the manifest Lorentz invariance
is to produce a non-polynomial action, but this difficults the quantization.
An alternative procedure is to use the Hamiltonian formalism in which, after
the introduction of an infinite set of fields (this idea was first used to
analyze chiral bosons \cite{W,CWY}) it is possible to get suitable duality
conditions \cite{BK}. The corresponding action, in ten dimensions,
containing an infinite number of fields, is manifestly invariant under
electromagnetic duality transformation. After compactification to four
dimensions it results into a local Maxwell action with electric and magnetic
sources \cite{NB}. The study of the dimensional dependence of the
electromagnetic duality was carried out in \cite{w2} and the connection
between duality and bosonization was shown in \cite{w3}.

In the approach of the source-free Maxwell theory in an arbitrary background
geometry, the duality symmetry under general electric/magnetic field
rotations can be implemented in a non-local way, as was introduced by Deser
and Teitelboim (DT) \cite{DT}. If the manifest covariance is lost, we can
construct quadratic actions for these models. It can be proved that this
formulation is equivalent to the Schwarz-Sen one's \cite{GGRS} via path
integral formalism. In reference \cite{DGHT} it was introduced sources into
these non-manifestly invariant actions and the covariantization of this
procedure has been accomplished in \cite{MB}.

Khoudeir and Pantoja (KP) \cite{KP}, getting back the non-polynomial action
scheme, suggested a Lorentz invariant version of the SS model by using an
auxiliary time-like constant vector in the action. This vector in fact
violates the manifest covariance. Pasti, Sorokin and Tonin (PST) \cite{PST},
making good use of this idea, proposed a generalization of the DT, SS and KP
duality symmetric actions presenting the KP unit norm auxiliary vector as a
Lorentz frame vector field that can be related to the gravitation
interaction. The action constructed is manifestly covariant.

In this work we use the fermion-like formulation of E. Majorana \cite{EM,MRB}
for the Maxwell theory to propose an action that is invariant under
relativistic and duality transformations. Fermion-like formulation for the
electromagnetic theory has been studied extensively by Dvoeglazov \cite
{russo}, but their connections with duality has not been explored. The paper
is organized as follows: in section 2 we review the work of Majorana and
obtain a Dirac-like expression for the Maxwell equations in the absence of
sources. This femion-like formulation suggest us to explore the chiral
aspects in the theory. This is accomplished in section 3 where we show that
duality is a kind of chirality and in this context we propose our action
which is invariant under duality. In section 4 we demonstrate that the
Poincar\'e generators obeys the algebra on-shell. Finally, the conclusion and
final observations are in section 5.

\section{Fermion-like formulation}

%\section{Fermionic formulation}\setcounter{equation}{0}

The well known Maxwell's equations are 
\begin{equation}  \label{tresa}
{\bf \nabla} \cdot {\bf E} = \rho
\end{equation}
\begin{equation}  \label{seisb}
{\bf \nabla} \cdot {\bf B} = 0
\end{equation}
\begin{equation}  \label{tresb}
{\bf \nabla} \times {\bf B} \,-\, \frac{\partial {\bf E}}{\partial t} = {\bf %
j}
\end{equation}
\begin{equation}  \label{seisa}
{\bf \nabla} \times {\bf E} \,+\, \frac{\partial {\bf B}}{\partial t} =
0\;\;.
\end{equation}

Now we want to look for a fermion-like formulation of these equations based
on the work of Majorana \cite{MRB}. In his work Majorana enhanced the r\^ole
in electrodynamics of the complex quantity ${\bf F}={\bf E} - i{\bf B} $,
which was emphasized lately by Weinberg \cite{SW} and others \cite{outros}.
The main motivation for this construction is the well known fact that at
statistical level, the electric and magnetic fields ${\bf E}$ and ${\bf B}$
are connected (through the quantity ${\bf E}^2+{\bf B}^2$) to the local mean
number of photons. Hence, an expression for the probability quantum wave of
a photon can be given in terms of ${\bf E}$ and ${\bf B}$. It provides a
meaning different of the usual one, where the electromagnetic four-potential
is introduced.

The first step is to build elements and vectors that allow us to couple two
equations in only one. To accomplish this we note that equations (\ref{tresb}%
) and (\ref{seisa}), for ${\bf j}=0$, can be rewritten as 
\begin{equation}  \label{setea}
i\,\frac{\partial {\Bbb E}}{\partial t}\,=\,{\frac{1 }{i}}\,({\bf s} \cdot 
{\bf \nabla})\,i\,{\Bbb B}
\end{equation}
and 
\begin{equation}  \label{seteb}
i\,\frac{\partial (i\,{\Bbb B})}{\partial t}\,=\,{\frac{1 }{i}}\,({\bf s}
\cdot {\bf \nabla})\,{\Bbb E}
\end{equation}
where 
\begin{equation}
{\Bbb E}\,=\,\left( 
\begin{array}{c}
E_1 \\ 
E_2 \\ 
E_3
\end{array}
\right)\:\:\:\:\: \mbox{and}\:\:\:\:\: {\Bbb B}\,=\,\left( 
\begin{array}{c}
B_1 \\ 
B_2 \\ 
B_3
\end{array}
\right)\;\;.
\end{equation}

\noindent Note that we have defined ${\bf s}$ as three matrices $3\times 3$,
by %\begin{equation}
$(s^{i})_{jk} =-i\,\epsilon _{ijk}$, %\label{oito}
%\end{equation}
where $\epsilon _{ijk}$ is the Levi-Civita totally antisymmetric tensor
normalized so that $\epsilon _{123}=1$ and $(\,{\bf s} \cdot {\bf \nabla}\,) 
\,{\Bbb B} = (s_{i})^{jk}\,\partial_j B_k$.

With this definition the $s^i$ matrices have the explicit structure 
\begin{eqnarray}
s^1\,&=&\;\left( 
\begin{array}{ccc}
0 & 0 & 0 \\ 
0 & 0 & -i \\ 
0 & i & 0
\end{array}
\right)\:,\:\: s^2\,=\;\left( 
\begin{array}{ccc}
0 & 0 & i \\ 
0 & 0 & 0 \\ 
-i & 0 & 0
\end{array}
\right)\:,  \nonumber \\
& &\:\:\:\:\:\:\:\:\:\:\:\: s^3\,=\;\left( 
\begin{array}{ccc}
0 & -i & 0 \\ 
i & 0 & 0 \\ 
0 & 0 & 0
\end{array}
\right)\:,
\end{eqnarray}
which satisfy the angular-momentum algebra 
\begin{equation}
[s_i,s_j]_-\,=\,-\,i\,\epsilon_{ijk}\,s_k\;,\;\;(i,j,k\,=\,1,2,3)\;\;,
\end{equation}
and in this way the equation (\ref{setea}), for example, can be rewritten as 
\begin{equation}
i\,\frac{\partial E_i}{\partial t}\,=\,(s_i)^{jk} \cdot {\partial_j}\, B_k
\end{equation}

In order to express equations (\ref{setea}) and (\ref{seteb}) in a
fermion-like formulation we define two important $6 \times 6$ matrices, 
\begin{equation}
\Gamma_0\,=\,\left( 
\begin{array}{cc}
{\Bbb I} & 0 \\ 
0 & -{\Bbb I}
\end{array}
\right)\:\:\:\:\: \mbox{and}\:\:\:\:\: {\vec\Gamma}\,=\,\left( 
\begin{array}{cc}
0 & {\bf s} \\ 
-{\bf s} & 0
\end{array}
\right)\;\;.
\end{equation}

\noindent $\Gamma_0$ and $\vec\Gamma$ play the r\^ole of the Dirac gamma
matrices, although they do not obey the usual gamma anticommutation
relations. We can also introduce a matrix that is equivalent to Dirac's $%
\gamma_5$, 
\begin{equation}
\Gamma_5\,=\,\left( 
\begin{array}{cc}
0 & {\Bbb I} \\ 
{\Bbb I} & 0
\end{array}
\right)
\end{equation}

\noindent where ${\Bbb I}$ is the $3 \times 3$ identity matrix. One can
check that $\{\,\Gamma_5\,,\,\Gamma_\mu\,\}\,=\,0\,\,,\,\mu=0,\ldots,3$.

With these definitions equations (\ref{setea}) and (\ref{seteb}) can be put
in a more compact form 
\begin{eqnarray}  \label{dez}
& &i\,\frac{\partial \Psi}{\partial t}\,=\,{\frac{1 }{i}}\,({\Bbb S} \cdot 
{\bf \nabla})\,\Psi  \nonumber \\
&\Rightarrow& (\,\partial_t\,+\,{\Bbb S} \cdot {\bf \nabla}\,)\,\Psi\,=\,0\:,
\end{eqnarray}
where ${\Bbb S}$ is 
\begin{equation}
{\Bbb S}\,=\, \left( 
\begin{array}{cc}
0 & {\bf s} \\ 
{\bf s} & 0
\end{array}
\right)\:.
\end{equation}
Eq. (\ref{dez}) resembles the massless Dirac equation, and one can consider $%
\Psi$ as a (quantum) wave function for the photon of the type 
\begin{equation}
\Psi\,=\;\left( 
\begin{array}{c}
{\Bbb E} \\ 
i\,{\Bbb B}
\end{array}
\right)
\end{equation}
and 
\begin{equation}
\overline{\Psi}\,=\;\left( 
\begin{array}{cc}
{\Bbb E}^\dagger & \:\:\:i{\Bbb B}^\dagger
\end{array}
\right)
\end{equation}

\noindent where $\overline{\Psi}=\Psi^\dagger \,\Gamma_0$ is an analog of
the Hermitian conjugated definition. Notice that only ${\bf E}$ and ${\bf B}$
have physical meaning and we will use ${\bf E}^\dagger$ and ${\bf B}^\dagger$
as auxiliary fields. So, we can note that $\Psi^\dagger \, \Psi={\bf E}^2 + 
{\bf B}^2$, with ${\bf E}^\dagger = {\bf E}$ and ${\bf B}^\dagger = {\bf B}$%
. Now we can see the importance of the complex notation in the construction
of $\Psi$ which mimics a Dirac spinor.

In terms of the $\Gamma$ matrices we may rewrite (\ref{dez}) as 
\begin{eqnarray}  \label{treze}
i\,\Gamma_0\,\frac{\partial \Psi}{\partial t}\,&=&\,{\frac{1 }{i}}%
\,(\Gamma_0\,{\Bbb S} \cdot {\bf \nabla})\,\Psi  \nonumber \\
&=&\,{\frac{1 }{i}}\,({\vec\Gamma} \cdot {\bf \nabla})\,\Psi
\end{eqnarray}
\begin{equation}
\Rightarrow (\,\Gamma_0\,\partial_t\,+\,{\vec\Gamma} \cdot {\bf \nabla}%
\,)\,\Psi\,=\,0
\end{equation}

\noindent or compactly 
\begin{equation}  \label{dezessete}
\Gamma^{\mu}\,\partial_\mu\,\Psi\,=\,0\;\;,
\end{equation}
though it is not manifestly covariant.

\section{Chirality and Duality transformations}

%\section{Chirality and Duality transformations}\setcounter{equation}{0}

We can also note that eq. (\ref{dezessete}) is invariant under the chiral
transformation 
\begin{equation}  \label{dezoito}
\Psi \rightarrow \Psi\,=\,e^{i\,\theta\,\Gamma_5}\,\Psi
\end{equation}
In terms of the fields ${\Bbb E}$ and ${\Bbb B}$ we have 
\begin{eqnarray}  \label{dezenove}
{\Bbb E}\,&=&\,cos \theta {\Bbb E}\,-\,sin \theta {\Bbb B}  \nonumber \\
{\Bbb B}\,&=&\,cos \theta {\Bbb B}\,+\,sin \theta {\Bbb E}\;\;,
\end{eqnarray}
which is a rotation and, particularly for $\theta=-\pi/2$ we recover the
duality transformation.

This invariance can also be verified in the following Lagrangian density 
\begin{equation}  \label{vintenove}
{\cal L}\,=\,i\,\alpha\,\overline{\Psi}\,(\,\Gamma_0\,\partial_t\,+\,{\vec%
\Gamma} \cdot \nabla\,)\,\Psi\;\;,
\end{equation}

\noindent or compactly ${\cal L}=i\,\alpha\,\Psi\,\Gamma^{\mu }\,\partial
_{\mu }\Psi $, where the vectors $\Psi$ and $\Psi^{\dagger }$ must be taken
conveniently as independent. In order to provide the correct dimension of
the action in the natural units $(\hbar=c=1)$ we have introduced a parameter 
$\alpha$ whose dimension is $mass^{-1}$ and does not modify the equations of
motion.

In terms of ${\Bbb E}$ and ${\Bbb B}$ eq. (\ref{vintenove}) is given by 
\begin{equation}  \label{vintenoveb}
{\cal L}\,=\,\alpha\,(\,i \, {\Bbb E}^\dagger \dot{{\Bbb E}} \,+\,i\,{\Bbb B}%
^\dagger \dot{{\Bbb B}} \,-\, \,{\Bbb E}^\dagger\,{\bf s} \cdot {\bf \nabla} 
{\Bbb B}\,+\,{\Bbb B}^\dagger\,{\bf s} \cdot {\bf \nabla} {\Bbb E}\,)\;\;,
\end{equation}
which is first-order in time derivative and consequently has a Hamiltonian
formulation. The physical quantities of the theory are real and the main
r\^ole of the auxiliary fields is in the mathematical construction of the
Maxwell equations as we will see below.

The equations of motion for ${\bf E}^\dagger$ and ${\bf B}^\dagger$ can be
obtained from the Euler-Lagrange equations 
\begin{equation}
\frac{\partial {\cal L}}{\partial E_i} \,-\,\partial_t \frac{\partial {\cal L%
}}{\partial \dot{E}_i} \,-\, \partial_j\, \frac{\partial {\cal L}}{\partial
(\partial_j E_i)}\,=\,0  \nonumber \\
\end{equation}
and 
\begin{equation}
\frac{\partial {\cal L}}{\partial B_i} \,-\,\partial_t \frac{\partial {\cal L%
}}{\partial \dot{B}_i} \,-\, \partial_j\, \frac{\partial {\cal L}}{\partial
(\partial_j B_i)}\,=\,0\;\;,
\end{equation}

\noindent with the following result, 
\begin{equation}
-\,i\,\frac{\partial {\Bbb E^\dagger}}{\partial t}\,=\,{\frac{1 }{i}}\,({\bf %
s} \cdot {\bf \nabla})\,i\, {\Bbb B^\dagger}
\end{equation}
and 
\begin{equation}
-\,i\,\frac{\partial (i\,{\Bbb B^\dagger})}{\partial t}\,=\,{\frac{1 }{i}}\,(%
{\bf s} \cdot {\bf \nabla})\, {\Bbb E^\dagger}\;\;,
\end{equation}
which are the Hermitian conjugated of (\ref{setea}) and (\ref{seteb})
respectively. The equations of motion for ${\bf E}$ and ${\bf B}$ are
obtained from 
\begin{equation}
\frac{\partial {\cal L}}{\partial E_i^{\dagger }} \,=\,0 \:\:\:\:\: %
\mbox{and} \:\:\:\:\: \frac{\partial {\cal L}}{\partial B_i^{\dagger }}
\,=\,0\;\;,
\end{equation}
and as a result we have equations (\ref{setea}) and (\ref{seteb}).

Equations (\ref{tresa}) and (\ref{seisb}) for $\rho=0$, comes from the fact
that ${\bf E}$ and ${\bf B}$ are orthogonal to the propagation vector,
namely 
\begin{equation}
{\bf k} \cdot {\bf E}\,=\,0\:\:\:\:\:\:\: \mbox{and}\:\:\:\:\:\:\: {\bf k}
\cdot {\bf B}\,=\,0\;\;.
\end{equation}

\bigskip

\noindent {\bf The mass term.} It is well known that a term like $\overline{%
\Psi }\,\Psi $ breaks the chiral invariance. In this formulation it is given
by 
\begin{equation}
\overline{\Psi }\,\Psi \,=\,{\Bbb E}^{\dagger }\cdot {\Bbb E}\,-\,{\Bbb B}%
^{\dagger }\cdot {\Bbb B}\;\;.
\end{equation}
which resembles the Maxwell Lagrangian density (with ${\bf E^\dagger} ={\bf E%
},{\bf B^\dagger}={\bf B}$) and highlights the connection between chiral
symmetry and duality. A mass term in the Dirac-form is not invariant under
chiral transformation and the Maxwell Lagrangian is not invariant under
duality transformation.

\section{The generators of the Poincar\'e algebra}

%\section{The generators of the Poincar\'e algebra}\setcounter{equation}{0}

Notice that, as the Schwarz-Sen model, eq. (\ref{vintenoveb}) is not a
Lorentz scalar. However, it is our task now to demonstrate that the theory
is relativistically invariant as it obeys the Poincar\'e algebra.

We have computed the generators of the Poincar\'{e} algebra in the usual
way, that is, by using the energy-momentum tensor as 
\begin{equation}
\Theta_{\mu\nu}\,=\,\frac{\partial {\cal L}}{\partial (\partial^\mu\,\phi_i)}%
\, \frac{\partial \phi_i}{\partial x^\nu}\,-\,g_{\mu\nu}\,{\cal L}
\end{equation}
where $\phi_i$ are the basic fields $E_i(E_i^\dagger)$ and $B_i(B_i^\dagger)$%
. Consequently the linear and angular momenta are given by, 
\begin{eqnarray}
H &=&P_{0}\,=\,\int\,d^3\,x\,\Theta_{00}  \nonumber \\
&=& \,\alpha\,\int \,d^{3}\,x\,[\,{\Bbb E}^{\dagger }\,{\bf s}\cdot {\bf %
\nabla}\,{\Bbb B}\,-\,{\Bbb B}^{\dagger }\,{\bf s}\cdot {\bf \nabla}\,{\Bbb E%
}\,]  \nonumber \\
P_{k} &=&\int\,d^3\,x\,\Theta_{0k}  \nonumber \\
&=&i\,\alpha\,\int \,d^{3}\,x\,[\,{\Bbb E}^{\dagger }\,\partial _{k}\,{\Bbb E%
}\,+\,{\Bbb B}^{\dagger }\,\partial _{k}\,{\Bbb B}\,]\;\;,
\end{eqnarray}
and 
\begin{eqnarray}
M_{\mu \nu } &=&i\,\alpha\,\int d^{3}\,x\,[\,x_{\mu }\,(\,{\Bbb E}^{\dagger
}\,\partial _{\nu }\,{\Bbb E}\,+\,{\Bbb B}^{\dagger }\,\partial _{\nu }\,%
{\Bbb B}\,)  \nonumber \\
&-&x_{\nu }\,(\,{\Bbb E}^{\dagger }\,\partial _{\mu }\,{\Bbb E}\,+\,{\Bbb B}%
^{\dagger }\,\partial _{\mu }\,{\Bbb B}\,)\,]\;\;,
\end{eqnarray}
where $\mu ,\nu \,=\,0,\ldots ,3$ and $k\,=\,1,\ldots ,3$. Finally it can be
shown in a straightforward calculation that they obey the Poincar\'e algebra
on-shell, i.e., 
\begin{eqnarray}
\{\,P_{k}\,,\,P_{0}\,\} &=&0  \nonumber \\
\{\,M_{\mu \nu }\,,\,P_{0}\,\} &=&0  \nonumber \\
\{\,M_{\mu \nu }\,,\,P_{k}\,\} &=&g_{k\nu }\,P_{\mu }\,-\,g_{k\mu }\,P_{\nu }
\nonumber \\
\{\,M_{\mu 0}\,,\,P_{k}\,\} &=&g_{k\mu }\,P_{0}  \nonumber \\
\{\,M_{\mu \nu }\,,\,M_{\lambda \sigma }\,\} &=&g_{\nu \lambda }\,M_{\mu
\sigma }\,-\,g_{\mu \sigma }\,M_{\nu \lambda }\,  \nonumber \\
&-&\,g_{\nu \sigma }\,M_{\mu \lambda }\,-\,g_{\mu \lambda }\,M_{\nu \sigma
}\;\;.
\end{eqnarray}
So, we have demonstrated that the action in eq. (\ref{vintenoveb}) describes
a theory that is relativistically invariant on-shell.

\section{Conclusions}

We have analyzed an already known fermion-like formulation for the
electromagnetic theory and proposed a corresponding Lagrangian invariant
under duality transformation. This invariance can be seen as a particular
case of chiral transformation in the fermion-like formulation. We have
carried out our analysis in order to verify the Poincar\'{e} invariance of
this formulation, and have found that, although not being manifestly
covariant, it is relativistically invariant on-shell. We also have compared
the chiral variant mass term with the Maxwell Lagrangian, which is duality
noninvariant. As a perspective for the future we intend to study the duality
symmetry in others formulations of the electromagnetism.

\section{Acknowledgments}

The authors would like to thank the financial support of Funda\c{c}\~{a}o de
Amparo \`{a} Pesquisa do Estado de S\~{a}o Paulo (FAPESP), a brazilian
research agency, professors A. S. Dutra and C. Wotzasek for valuable
discussions and professor V. Dvoeglazov for helpful comments.

%\end{\large}

\end{document}